\documentclass[11pt]{article}
\usepackage{}
\usepackage{amsfonts}
\usepackage{bbm}
\usepackage{bm}
\usepackage{mathrsfs}
\usepackage{color}
\usepackage{geometry}             % See geometry.pdf to learn the layout options. There are lots.
\usepackage{graphicx}
\usepackage{amssymb}
\usepackage{amsmath}
\usepackage{cite}
\usepackage{comment}

\usepackage{ulem}

\usepackage[hyperindex=true,
          pdfstartview=FitH,
          bookmarksnumbered=true,
          bookmarksopen=true,
          citecolor=blue,
          linkcolor=blue,
          colorlinks=true,
          urlcolor=rossoCP3,
          pdfborder=001,
          unicode]{hyperref}

\definecolor{rossoCP3}{cmyk}{0,.88,.77,.40}

\newcommand{\rd}[1]{\mathrm{d}#1}

\newcommand{\re}{\mathrm{e}}
\newcommand{\la}{\langle}
\newcommand{\ra}{\rangle}
\newcommand{\g}{\mathfrak{g}}

\newcommand{\sech}{\mathrm{sech}}
\newcommand{\pf}[2]{\frac{\partial #1}{\partial #2}}

\usepackage{xcolor}  % Required for custom colors
\definecolor{mygreen}{RGB}{44,85,17}
\definecolor{myblue}{RGB}{34,31,217}
\definecolor{mybrown}{RGB}{194,164,113}
\definecolor{myred}{RGB}{255,66,56}
\definecolor{mypurple}{RGB}{200,36,176}
\allowdisplaybreaks

\newcommand{\blue}[1]{\textcolor{blue}{#1}}
\renewcommand{\blue}[1]{\textcolor{black}{#1}}  
\begin{document}

\title{Covariant transport equation and \\
gravito-conductivity\\
in generic stationary spacetimes}
\author{Song Liu${}^1$, Xin Hao${}^2$, Shaofan Liu${}^3$,
and Liu Zhao${}^1$\thanks{Correspondence author.}
\vspace{1pt}\\
\small ${}^1$School of Physics, Nankai University, Tianjin 300071, China\\
\small ${}^2$School of Physics, Hebei Normal University, Shijiazhuang 050024, China\\
\small ${}^3$Department of Physics, University of Houston, Houston, TX 77204, USA\\
{\small {email}:
\href{mailto:2120190124@mail.nankai.edu.cn}
{2120190124@mail.nankai.edu.cn}} \\
{\small \href{mailto:xhao@hebtu.edu.cn}
{xhao@hebtu.edu.cn},
\href{mailto:sliu74@cougarnet.uh.edu}
{sliu74@cougarnet.uh.edu}} \\
{\small and
\href{mailto:lzhao@nankai.edu.cn}{lzhao@nankai.edu.cn}}
}
\date{}
\maketitle
\begin{abstract}
  We find a near detailed balance solution to the relativistic Boltzmann equation
  under the relaxation time approximation with a collision term 
  which differs from the Anderson-Witting model and is dependent on the 
  stationary observer. Using this new 
  solution, we construct an explicit covariant transport equation for the particle flux 
  in response to the generalized temperature and chemical potential gradients
  in generic stationary spacetimes, with the transport tensors characterized 
  by some integral functions in the chemical potential and the relativistic coldness. 
  To illustrate the application of the transport equation,
  we study probe systems in Rindler and Kerr spacetimes
  and analyze the asymptotic properties of the gravito-conductivity 
  tensor in the near horizon limit. It turns out that both the longitudinal and 
  lateral parts (if present) of the gravito-conductivity tend to be divergent 
  in the near horizon limit. In the weak field limit, our results coincide with 
  the non-relativistic gravitational transport equation which follows from 
  the direct application of the Drude model. 
\end{abstract}

%\newpage

\section{Introduction}

Transport equations for macroscopic systems under the influence of
external fields are critical in understanding non-equilibrium processes.
For electron gas in solid state physics, the external fields can be electromagnetic field
and/or temperature gradient. For compact stars and accretions around black holes, the
relativistic gravitational field must be taken into consideration.
In the recent years, there has been an increasing interest
in studying the evolution of kinetic gases in the vicinity of black holes.
On the one hand, the macroscopic properties of near horizon systems are likely to
be connected with thermodynamic effects of gravity.
On the other hand, the rotating plasma surrounding supermassive black holes 
has been revealed by Event Horizon Telescope 
Collaboration \cite{EventHorizonTelescope:2019dse}. Motivated by these 
astrophysical scenarios, it is necessary to dwell on some more details
of the transport phenomena in the strong gravity regime.

Kinetic theory has long been used for constructing transport equations. 
In the early stages of the universe and in the near horizon region, 
gravity is so strong that the relativistic effect becomes important. In this situation,
a relativistically covariant approach is needed. Such a systematic approach is known as the
relativistic kinetic theory (RKT) and could be dated back to 1911 
\cite{doi:10.1002/andp.19113390503}. The modern geometrical description of 
specially relativistic kinetic theory is formulated in
\cite{Synge:1934zzb}, which is fundamental for the 
generalization to curved spacetime backgrounds
\cite{PhysRev.122.1342,Chernikov1963,doi:10.1063/1.1704047}.
Renewed interests in this area arise following Refs.\cite{Heinz:1983nx,Bazow:2015dha,
Rioseco:2016jwc}, where RKT has been respectively applied
to the study of the quark-gluon plasma and heavy ion collisions 
\cite{Guo-Qiang:1996,Miake:2005,Mendoza:2014,Greif:2014oia,Simeoni:2019,
ChunShen:2022,Elze:1987ii,Danielewicz:1991dh,Bertsch:1988,TMEP:2022xjg},
the early Universe \cite{Hannestad:1995rs,Hu:1995em,Uzan:1998mc,Birrell:2014gea,
Husdal:2016pfd,Adshead:2016xxj,Sasankan:2019oee,Pordeus-da-Silva:2019bak,
Pitrou:2019hqg,Sarbach2022} and astrophysics\cite{Molnar:2020nfd,Deng:2021twz,
Nishikawa:2020rwe,Mach:2022mrn}, particularly accretion around black 
holes\cite{Rioseco:2016jwc,Mach:2022mrn,Rioseco:2017b,Gabarrete:2021}.
The covariant formalism also led to the development of relativistic 
thermodynamics \cite{Israel:1976tn,Israel:1979wp}. 

The aim of the present work is to provide a covariant formulation for transport phenomena
in curved spacetimes. There have been studies about the gravitational effects on
the transport of conserved charge, heat and momentum \cite{Anderdon:1973,
Cercignani:2002,deGroot}. The major difference between
the present work and the previous ones lies in that we emphasize the importance of
the observer, and that our collision model is independent of Landau frame. 
The reason for stressing the important role of observers lies as follows. 
As we have pointed out in \cite{Hao:2021ifw}, in relativistic physics, 
physical laws are independent of the choice of coordinate system, whilst the 
values of physical observables are observer dependent. 
In the description of relativistic transport phenomena, the transport equation 
encodes the underlying physical law which needs to be 
described in a covariant formalism. On the other hand, 
the values of the relevant transport coefficients --- although also need to be 
tensorial objects --- describe the phenomenological properties of 
concrete systems which depend on the choice of observer. As will be seen in the main 
texts below, the proper velocity of the stationary \blue{observer} is crucial when 
modeling the collision term and interpreting the kinetic tensor.

Another difference of the present work with previous ones is that the lateral 
transport in our framework can be
clearly separated from the longitudinal transport. One may think that
the lateral effects are negligibly small as compared to
the longitudinal effects. This may be a truth in some cases, but not always.
For instance, around rapidly spinning black holes, the lateral  effects are 
at least equally as important as the longitudinal effects.

For simplicity, we will work on the assumption that
the gaseous (or fluid) system is consisted of massive neutral particles 
and is regarded as a probe system.

\section{GEM field and the gravitational Hall coefficient
under post-Newtonian approximation}

In this section, we work within the post-Newtonian limit,
i.e. when the gravitational field is weak, and the motion
of the source is slow but still taken into consideration.
In this limit, the gravitational field can be described by
small perturbations $h_{\mu\nu}$ around Minkowski metric.
When all the nonlinear terms in $h_{\mu\nu}$ are neglected, there will be an analogy between
Einstein equations under de Donder gauge and the Maxwell equation under \blue{Lorenz} gauge.
In 4-dimensional spacetimes, 
such an analogy is known as the linear gravito-electromagnetism (GEM)
\cite{Thorne,Mashhoon:2000he,DeWitt:1966yi,Dessler:1968zz}. The solution
of the linearized Einstein equation suggests that, up to $\mathcal O(c^{-2})$,
we can write the metric as
\begin{align}
\mathrm ds^2 = -c^2 \Big(1+\frac{2\phi}{c^2}\Big) \mathrm dt^2
+ \frac{4}{c}\bm A \cdot \rd \bm x\,
\rd t + \Big(1-\frac{2\phi}{c^2}\Big)
\delta_{ij} \mathrm dx^i \mathrm dx^j,
\end{align}
where $i,j=1,2,3$, $\phi = h_{00}/2c^2$ is the scalar potential and $A_{i} = h_{0i}/2c^2$
is the vector potential. For slowly varying
source, the gravito-electric (GE) field and the gravito-magnetic (GM) field
become $\bm E^{(g)} = -\nabla \phi$ and $\bm B^{(g)} = \nabla \times \bm A$,
which satisfy a set of equations analogous to the Maxwell equations in 
3-vector formulation.

Another important aspect of the linear GEM framework lies in the equation of
motion of a test particle. In non-relativistic limit, the proper velocity
of the test particle is approximately $u^\alpha \approx (c , \bm v)$,
and up to ${\mathcal O}(c^{-1})$ the geodesic equation in terms of GE and GM
reduces to Lorentz-force-like equation
\begin{align}
\frac{\rd \bm v}{\rd t} = \bm{E}^{(g)} +  2\,\frac{\bm{v}}{c} \times\bm{B}^{(g)},
\label{Lorentz}
\end{align}
\begin{comment}
{\color{blue}
\begin{align*}
\frac{\rd u^\alpha}{\rd s} + \Gamma^\alpha{}_{\mu\nu} u^\mu u^\nu = 0 \quad
&\dashrightarrow \quad \frac{\rd u^i}{\rd t}
+ \left(c^2 \Gamma^i{}_{00}
+ 2c\,v^j\Gamma^i{}_{j0}
+ v^j v^k \Gamma^i{}_{jk}\right) = 0 \\
\frac{\rd \bm v}{\rd t} = \bm{E}^{(Z)} +  2\,\frac{\bm{v}}{c}\times\bm{B}^{(Z)}
&\dashleftarrow \frac{\rd v^i}{\rd t} + \left[\partial^i \phi
+  \frac{2}{c} v^j\left(\partial_j A^{(g)}_k-\partial_k A^{(g)}_j\right)
\delta^{ki}\right] =0
\end{align*}
}
\end{comment}
as a simple application of eq.\eqref{Lorentz}, we can now
follow the {\it Drude model} of traditional Hall effect and introduce
a linear friction term $- \bm v/ \tau$ to the RHS, which is due to scattering
between particles,
\begin{comment}
{\color{blue}
\begin{align}
\frac{\rd \bm v}{\rd t} = \bm{E}^{(Z)} +  2\,\frac{\bm{v}}{c} \times\bm{B}^{(Z)} - \frac{\bm v}{ \tau},
\end{align}
}
\end{comment}
and the scattering time $\tau$ is the average time between collisions. When
the particle becomes equilibrated, we have roughly $\rd \bm v /\rd t = 0$. 
In this case, 
\begin{comment}
{\color{blue}
\begin{align*}
\bm v -  2 \tau\,\frac{\bm{v}}{c} \times\bm{B}^{(Z)} = \tau \bm{E}^{(Z)},
\end{align*}
}
\end{comment}
the particle flow $\bm j = n \bm v$ is related to external force $\bm E^{(g)}$ by
\begin{align}
\mathcal R^{(g)} \bm j -  \mathcal R^{(g)}_H \bm j \times\bm{B}^{(g)} =  \bm{E}^{(g)},
\label{Hall}
\end{align}
where $\mathcal R^{(g)} = 1/n\tau$ is the gravitational resistance and
$\mathcal R^{(g)}_H = 2/nc$ is the gravitational Hall coefficient for
neutral systems. This analogy is clear and instructive, but the  Drude model 
is highly dependent on the Lorentz-force-like
equation, therefore only applies to weak gravitational field. 

In the rest of this work, we will generalize the transport equation to 
generic stationary spacetime backgrounds in a fully covariant form 
and calculate the gravitational conductivity for neutral systems 
as an example case for applications.
\begin{comment}
finally the current density $\bm J = n \bm v$ is related to $\bm E^{(Z)}$ by
a matrix coefficient
\begin{align*}
\left(\delta_a{}^b + \tau \omega_a{}^b \right) J_b = - n\tau  E^{(Z)}_a
\end{align*}
or
\begin{align*}
J_a = -n\tau\left[\delta_a{}^b+
\left(1 - \tau^2 \omega_i{}^j \omega_j{}^i\right)^{-1}
\left(-\tau \omega_a{}^b + \tau^2 \omega_a{}^c \omega_c{}^b\right)\right] E^{(Z)}_b
\end{align*}
For example, for rotating system
\begin{align*}
\rd  s^2 = - \left(1-\frac{\omega^2r^2}{c^2}\right) c^2 \rd t^2 +
\rd r^2 + r^2 \rd \varphi^2 + \rd z^2
+ 2\omega r^2 \rd t \rd \varphi, \qquad
\mathcal B^\mu=\frac{\beta\left(\partial_t\right)^\mu}{\sqrt{1-\omega^2 r^2/c^2}},
\end{align*}
the only non-zero component of $\omega_{ab}$ is $\omega_{12} = -\omega_{21}= B^{(Z)}_{\,3}$, so
\begin{align*}
-\frac{ 1}{n\tau }\left( \begin{matrix}
	1&		\tau B^{(Z)}_{\,3} \\
	 -\tau B^{(Z)}_{\,3} &		1\\
\end{matrix} \right)\bm J = \bm E^{(Z)}
\end{align*}
or
\begin{align*}
\bm J= -\frac{n\tau }{1 + \tau^2 [ B^{(Z)}_{\,3}]^2 }\left( \begin{matrix}
	1&		- \tau B^{(Z)}_{\,3} \\
	\tau B^{(Z)}_{\,3} &		1\\
\end{matrix} \right)\bm E^{(Z)}
\end{align*}
Finally the gravitational Hall coefficient is
\begin{align*}
R^{(g)}_H = - \frac{E^{(Z)}_{\,2}}{J_1 B^{(Z)}_{\,3}} = \frac{1}{n}
\end{align*}
which is independent of the the scattering time
\end{comment}

\section{Relativistic Boltzmann equation and approximate solution}

Before proceeding, let us mention that, in ordinary electrodynamics, 
different observers should observe different electric ($\bm{E}$) and 
magnetic ($\bm{B}$) fields, which is known as the Faraday effect. Likewise, 
The GE and GM fields must also be observer dependent. The description presented 
in the last section did not show this observer dependence, because a specific
observer, i.e. the static observer in the Minkowski background spacetime, is
implicitly adopted. 

In the following, we will try to establish a fully covariant formalism for 
the relativistic transport equations. For this purpose, we need to 
bring back the explicit observer dependence. It turns out that the 
stationary observer is best suited for our investigation. 
Therefore we shall assume that the spacetime is stationary 
(i.e. there exists a timelike Killing vector field). 

From now on, we will be working in natural units with $c=h=k_{\rm B}=1$. 
We introduce a generic observer $\mathcal{O}$ with 
proper velocity $Z^\mu$ (with $Z^\mu Z_\mu=-1$) and the observer dependence of the
GE and GM fields are best illustrated by their expressions in terms of $Z^\mu$, i.e.  
\begin{align}
E^{(g)}_{\mu} = - Z^\nu \nabla_\nu Z_\mu,
\qquad B^{(g)}_{\mu\nu} = \frac{1}{2}\nabla_{[\alpha} Z_{\beta]}
\Delta^\alpha{}_\mu \Delta^\beta{}_\nu,
\label{GEGB}
\end{align}
\begin{comment}
\begin{align}
E^{(g)}_{\mu} = -\frac{1}{2}
Z^\nu \nabla_\nu Z_\mu,
\qquad B^{(g)}_{\mu\nu} = c\,\nabla_{[\alpha} Z_{\beta]}
\Delta^\alpha{}_\mu \Delta^\beta{}_\nu = -c\, \omega_{\mu\nu},
\label{GEGB}
\end{align}
\end{comment}
where $\Delta_{\mu\nu} = g_{\mu\nu} + Z_\mu Z_\mu$ is the normal projection tensor 
associated with the observer $\mathcal{O}$. It should be mentioned that the covariant
GE field $E^{(g)}_{\mu}$ is proportional to the proper acceleration of the 
observer $\mathcal{O}$. The GM field represented in vector form can be introduced as
\begin{align*}
B^{(g)\mu} = \blue{-}
\Delta^{\mu}{}_\nu\varepsilon^{\nu\alpha\beta} B^{(g)}_{\alpha\beta},
\end{align*}
where $\varepsilon^{\nu\alpha\beta}$ is the Levi-Civita tensor.

Now let us proceed with a brief review of the relativistic kinetic theory for a system
composed of neutral particles moving in curved spacetime. The macroscopic state of the
system is described by the particle number flow $N^\mu$, the energy momentum 
tensor $T^{\mu\nu}$ and  the entropy current $S^\mu$. These quantities are 
primary in the fluid description of macroscopic systems
\cite{Israel}. 
In the absence of external field besides gravity, one has
\begin{align}
\nabla_\mu N^\mu = 0, \qquad \nabla_\mu T^{\mu\nu} = 0,
\qquad \nabla_\mu S^\mu \geqslant 0.    \label{NTS}
\end{align}
The expressions for $N^\mu$, $T^{\mu\nu}$ and $S^\mu$ in terms of
other macroscopic observables such as
temperature $T$, chemical potential $\mu$ and entropy density $s$
are called constitutive relations. It is important to keep in
mind that $T$, $\mu$, $s$ and therefore
the constitutive relations are observer dependent, but the above 
primary macroscopic quantities are not, since they have microscopic
definitions. In standard relativistic kinetic theory, the primary macroscopic
quantities are expressed as integrations over one-particle distribution
function (1PDF),
\begin{align}
&N^\mu=  \int \bm\varpi p^\mu f, \qquad
T^{\mu\nu}= \int\bm\varpi p^\mu p^\nu f,   \\
&S^\mu = - \int \bm\varpi p^\mu
\left[f \log f - \varsigma^{-1}
\left(1 + \varsigma f \right)
\log \left(1 + \varsigma f\right)\right], 
\end{align}
where $\varsigma=0,+1,-1$ respectively corresponds to 
the non-degenerate, bosonic and fermionic cases, 
$\bm\varpi = \g \frac{\sqrt{g}}{p_0} (\rd p)^d$ is the invariant volume element 
in the momentum space with $\g$ being the intrinsic degeneracy factor
for individual particles in the system, and
the spacetime dimension is chosen to be $(d+1)$. The above
definitions are valid for systems away from equilibrium. Therefore,
in kinetic theory, we can ask whether the equilibrium state
can be achieved in curved spacetime, and find the constraints for
the spacetime metric to embed an equilibrium system. Since the primary
macroscopic quantities are totally determined by the 1PDF,
the relativistic Boltzmann equation (RBE) is then of fundamental
importance
\begin{align}
\mathscr{L_{\mathcal H}} f= \mathcal C(f), \label{BE}
\end{align}
where $\displaystyle\mathscr{L_{\mathcal H}} = p^\mu\pf{}{x^\mu}
-\Gamma^\mu{}_{\alpha\beta}p^\alpha
p^\beta \pf{}{p^\mu}$ is the Liouville vector field on the tangent bundle of the spacetime.

There are two-fold complexities in the RBE, which make it challenging 
to find an exact solution. On the left hand side, the spacetime geometry
is encoded in the Liouville vector field, however the spacetime geometry itself
can only be determined by solving the Einstein equation with the 
energy-momentum tensor acting as the source, but to define the 
energy-momentum tensor one needs the 1PDF which is the sought-for solution of the 
RBE. Therefore, the RBE and Einstein equation together constitute a coupled 
system of differential-integral equations which is extremely difficult to solve. 
On the right hand side, the exact expression for the collision integral 
$\mathcal C(f)$ requires the knowledge of precise form of the transition matrix
for two particle scattering. 

Fortunately, most of the difficulties can be overcome by adopting the probe 
limit and by making certain simplifications for the collision integral by assuming
certain symmetries. The most commonly used assumption is that 
the transition matrix obeys time
reversal symmetry. Under this assumption, the vanishing of entropy production
rate indicates that $\mathcal C(f) = 0$, which is known as the
detailed balance condition. As a result, the RBE reduces to the Liouville equation,
and the solution is the detailed balance distribution
\begin{align}
\bar f=
%\frac{\g}{h_{\mathrm{P}}^d}
\frac{1}{\re^{\bar \alpha- \bar{\mathcal B}_\mu p^\mu} - \varsigma},
\label{fbar}
\end{align}
where $\bar \alpha$ and $\bar{\mathcal B}^\mu$ are independent of momentum,
$\bar\alpha$ is a constant scalar in spacetime
manifold and $\bar{\mathcal B}^\mu$ is a Killing vector field. Please keep in mind 
that quantities with an over bar are evaluated under detailed balance.
On account of the energy conditions for $T^{\mu\nu}$,
we only study the fluid system where $\bar{\mathcal B}^\mu$
is timelike. 

Although the primary tensor objects $N^\mu, T^{\mu\nu}, S^\mu$ are observer independent, 
their values under detailed balance are best expressed in terms of the proper
velocity of the {\it comoving observer}. Let us recall that an observer whose proper
velocity is proportional to the timelike Killing vector field is stationary, 
while an observer whose proper velocity is a timelike eigenvector field of $T^{\mu\nu}$
is comoving. Since we have already assumed that $\bar{\mathcal B}^\mu$ is timelike 
and Killing, we can determine the proper velocity $Z^\mu$ of the stationary observer 
via $\bar{\mathcal B}^\mu = \bar \beta Z^\mu$ and $Z^\mu Z_\mu=-1$. 
In terms of $Z^\mu$ and its 
corresponding projection tensor $\Delta_{\mu\nu}$, 
the primary tensor objects for a system consisted of massive particles of 
mass $m$ under detailed balance are evaluated to be
\begin{align}
&\bar{N}^\mu = \g m^d \mathcal A_{d-1} \bar J_{d-1,1}\, Z^\mu, \label{Ne} \\
&\bar T^{\mu\nu} = \g m^{d+1} \mathcal A_{d-1} 
\left(\bar J_{d-1,2} \,Z^\mu Z^\nu +\frac{1}{d} \bar J_{d+1,0}
\, \Delta^{\mu\nu} \right), \label{Tee} \\
&\bar S^\mu = \g m^d  \mathcal A_{d-1} 
\left(\bar\alpha \bar J_{d-1,1} + \bar\zeta \bar J_{d-1,2}
+\frac{\bar\zeta}{d} \bar J_{d+1,0}\right) Z^\mu , \label{Se}
\end{align}
where $\mathcal A_{d-1}$ is the area of the $(d-1)$-dimensional unit sphere, and 
$\bar  J_{nm}$ is the following function in $(\bar\alpha,\bar\zeta)$,
wherein $\bar\zeta = \bar\beta m$ is known as the relativistic coldness,
\begin{align*}
\bar  J_{nm} =
J_{nm}(\bar \alpha, \bar \zeta,\varsigma) = \int_0^\infty\frac{\sinh^n \vartheta
\cosh^m\vartheta}{\re^{\bar \alpha+\bar\zeta \cosh \vartheta} - \varsigma}
\rd \vartheta.
\end{align*}
As usual, the parameters $\bar\alpha,\bar\beta$ are related to the chemical 
potential $\bar\mu$ and temperature $\bar T$ measured by the comoving observer via
\[
\bar\alpha=-\bar\mu/\bar T,\qquad \bar\beta=1/\bar T.
\]

A key feature hidden in the above results is that 
$\bar N^\mu$, $\bar S^\mu $ and the eigenvector of $ \bar T^{\mu\nu}$ are collinear,
and the unique timelike eigenvector of $\bar T^{\mu\nu}$ coincides with
$Z^\mu$. Therefore, the stationary observer in this case is automatically comoving. 
Please be reminded that this feature is present only in the presence of 
detailed balance. 
%That is why the co-moving observer for a system in detailed balance
%is naturally defined as $Z^\mu$, and it is hidden in all the phenomenological
%thermodynamics relations we are familiar with. 
For systems away from detailed balance, there is no first principle 
definition for comoving observer. Therefore, in the following, we will
only consider stationary observers, and the results will be presented in a 
covariant way. 

In the eyes of a stationary observer, the 
temperature $\bar T$ and the chemical potential $\bar \mu$ under detailed 
balance satisfy
\[
\mathcal D_\nu \bar T \equiv \nabla_\nu \bar T
+\bar T \, Z^\sigma\nabla_\sigma Z_\nu = 0,
\qquad
\mathcal D_\nu \bar \mu \equiv \nabla_\nu \bar \mu
+\bar \mu \, Z^\sigma\nabla_\sigma Z_\nu = 0.
\]
We see that due to the Tolman-Ehrenfest effect and
the Klein effect, chemical potential gradient
and the temperature gradient alone can be
no longer viewed as thermodynamic forces in
gravitation field. In the above modification, we have
defined the {\it generalized temperature gradient} $\mathcal D_\nu T$
and the {\it generalized chemical potential gradient} $\mathcal D_\nu \mu$
which are responsible for thermodynamic forces.

Now we turn on external field by
setting $\mathcal D_\nu T, \mathcal D_\nu \mu \neq 0$,
while still taking the probe limit and keeping the spacetime metric fixed.
In this case, $\bar \alpha, \bar \beta$ will be replaced by $\alpha, \beta$ 
which are slightly different from the former, i.e. 
$\bar \alpha \to \alpha = \bar \alpha + \delta \alpha$, 
$\bar \beta \to  \beta = \bar \beta + \delta \beta$. In particular, 
there is no reason to take $\alpha$ as a constant. 
As a result, $\mathcal B^\mu\equiv \beta Z^\mu$ is no longer Killing,
although $Z^\mu$ remains to be the proper velocity of the stationary observer. 

Since we are now considering cases away from detailed balance, the 1PDF $f$ 
should differ from $\bar f$. If the deviation from detailed balance is not too
severe, we can reasonably assume that the zeroth order approximation to the 1PDF
takes the form
\begin{align}
f^{(0)}=
%\frac{\g}{h_{\mathrm{P}}^d}
\frac{1}{\re^{ \alpha- \mathcal B_\mu p^\mu} - \varsigma}
\label{f0}
\end{align}
and that the deviation of the full 1PDF $f$ from $f^{(0)}$ can be treated as a
perturbative modification. 

%Of course $f^{(0)}$ solves neither the Liouville equation nor the full RBE. 
In order to obtain a better approximate solution to the full RBE, 
one needs to simplify the collision integral by replacing the integral with some
algebraic expression while maintaining its basic symmetries. 
This kind of simplification is known as relaxation time approximation and its 
concrete realizations are known as collision models. In non-relativistic framework, 
the most widely known collision model is the
Bhatnagar-Gross-Krook (BGK) model \cite{BGK1954} which maintains the 
conservation of mass, momentum and energy. The direct application of BGK model 
in the relativistic case is called Marle-BGK model \cite{Marle1965} which 
seems to be oversimplified, since the relaxation
time becomes unbounded in ultra-relativistic limit. This problem was pointed out by
Anderson and Witting who successively proposed an alternative, adapted collision 
model which is referred to as Anderson-Witting (AW) model \cite{Anderdon:1973}. In AW model, 
Landau frame must be preassigned to guarantee the conservation laws for 
particle flow and energy-momentum tensor. However, in the present work, 
when external field is turned on, the conservation laws
becomes less important, instead, the generalized chemical potential and 
temperature gradients must be taken into account. In this case, 
the stationary observer plays a prominent role.
On the one hand, $Z^\mu$ is used to define $\mathcal D_\nu T$ 
and $\mathcal D_\nu \mu$. On the other hand,
when back reaction is neglected, $Z^\mu$ is independent of the perturbation. 
Following the above considerations, we propose the following 
collision model
\begin{align}
\mathscr L_{\mathcal H} f = - \frac{\varepsilon}{\tau} (f-f^{(0)})
=  \frac{Z_\mu p^\mu}{\tau} (f-f^{(0)}),
\label{Cxp}
\end{align}
where $\varepsilon=-Z_\mu p^\mu$ represents the energy of a single particle 
measured by the stationary observer, and $\tau$ is the time scale for the 
system to restore detailed balance. 
As the external fields $\mathcal D_\nu T$ and $\mathcal D_\nu \mu$ are
turned on, $Z^\mu$ does not necessarily coincide with the proper velocity of 
the fluid element in the Landau frame,
therefore, this collision model is different from AW model.

The desired solution for eq.\eqref{Cxp} is found to be
$f = f^{(0)} + \delta f$, where the modification $\delta f$ arises
due to the presence of the generalized temperature gradient and generalized chemical
potential gradient. Up to $\mathcal O(\tau^2)$, the explicit expression
for $\delta f$ reads
\begin{align}
\delta f
= \frac{\tau}{\varepsilon}\frac{\partial f^{(0)}}{\partial \varepsilon}
p^\mu\left(\delta_\mu{}^\nu + \frac{\tau}{\varepsilon}
\Delta^{\alpha}{}_\mu \nabla_{[\alpha}p_{\beta]}\,\Delta^{\beta\nu} \right)
\left(\mathcal D_\nu \mu
+ \frac{\varepsilon-\mu}{T} \mathcal D_\nu T\right).
\label{deltaf}
\end{align}
This equation is one of our main results and the recursive procedure for obtaining 
this approximate solution is given in appendix \ref{A}.

\section{Transport equations and gravito-conductivity}

The approximate solution to the RBE obtained in the last section allows us to study
various transport equations in a covariant manner. To illustrate such applications,
we will now take the particle number flow $N^\mu$ as an example and leave the 
detailed study about the transport behaviors for $T^{\mu\nu}$ and $S^\mu$ 
for later works.

According to the definition \eqref{NTS} for the particle number flow, 
we have 
\begin{align*}
N^\mu= \left[N^{(0)}\right]^\mu + \delta N^\mu =  \int \bm\varpi p^\mu f^{(0)}
+ \int \bm\varpi p^\mu \delta f.
\end{align*}
The first part of $N^\mu$, i.e. $\left[N^{(0)}\right]^\mu$, is collinear 
with $Z^\mu$, hence only a scalar density $n^{(0)}= -Z_\mu\left[N^{(0)}\right]^\mu$ 
is encoded in this part of the particle number flow, which reads 
\begin{align}
n^{(0)} = \g m^d \mathcal A_{d-1} J_{d-1,1},
\label{n0}
\end{align}
wherein $J_{d-1,1}$ is like $\bar J_{d-1,1}$ but with $\bar\alpha, \bar\zeta$ 
replaced by $\alpha, \zeta$, and therefore $n^{(0)}$ is slightly different 
from the particle number density $\bar n$ under the detailed balance condition. The second 
part of the particle number flow, i.e. $\delta N^\mu$, is not collinear 
with $Z^\mu$. One can project $\delta N^\mu$ onto parallel and orthogonal directions 
with respect to $Z^\mu$, where the parallel projection $\delta n\equiv 
-Z_\mu \delta N^\mu$ contributes a higher order correction to the particle 
number density $n= n^{(0)} +\delta n$, while the orthogonal projection 
gives rise to the following transport equation for the particle number, which reads
(see appendix \ref{B} for details)
\begin{comment}
\begin{align}
j^\mu = \Delta^\mu{}_\nu \delta N^\nu = - \frac{\tau}{m}
\frac{\mathcal A_{d-1}}{d} \frac{\g}{\lambda_C^d} \, \zeta
\left(\Delta^{\mu\nu} - \tau \omega^{\mu\nu}\right)
\Big[&\mathcal J_{d+1,-1}\mathcal D_\nu \mu  \nonumber \\
& +k_{\text{B}}(\alpha \mathcal J_{d+1,-1}
+\zeta \mathcal J_{d+1,0}) \mathcal D_\nu T\Big],
\end{align}
\end{comment}
\begin{align}
j^\mu \equiv \Delta^\mu{}_\nu \delta N^\nu = - \tau \g {m}^{d-1}
\Big(\sigma^{\mu\nu}\mathcal D_\nu \mu
+ \kappa^{\mu\nu} \mathcal D_\nu T\Big), \label{Transeq}
\end{align}
where the dimensionless factors $\sigma^{\mu\nu}$ and $\kappa^{\mu\nu}$ 
in the transport tensors are given by
\begin{align*}
\sigma^{\mu\nu} = \frac{\mathcal A_{d-1}}{d} \left(\Delta^{\mu\nu} 
+ 2\tau B^{(g)\mu\nu}\right)
\zeta \mathcal J_{d+1,-1}, \quad
\kappa^{\mu\nu} = \sigma^{\mu\nu} \left(\alpha 
+\zeta \frac{\mathcal J_{d+1,0}}{\mathcal J_{d+1,-1}}\right),
\end{align*}
in which $\mathcal J_{nm}$ is defined as
\begin{align*}
\mathcal J_{nm}(\alpha,\zeta,\varsigma) =
J_{nm}(\alpha,\zeta,\varsigma)+ \varsigma
\int_0^\infty\frac{\sinh^n \vartheta\cosh^m\vartheta}{\left(\re^{\alpha
+\zeta \cosh \vartheta} - \varsigma\right)^2}
\rd \vartheta.
\end{align*}

The covariant transport equation \eqref{Transeq} is valid for any probe system in 
any stationary spacetime, even in the strong field regime. Therefore, it can be used
for describing near horizon particle transport. Let us remark that
$\sigma^{\mu\nu}$ and $\kappa^{\mu\nu}$ differ from each other only by a scalar factor. 
Therefore, it suffices to consider only the {\it gravito-conductivity} 
characterized by $\sigma^{\mu\nu}$ in the following context (the actual 
gravito-conductivity differs from $\sigma^{\mu\nu}$ by a constant factor 
$- \tau \g {m}^{d-1}$, but we shall slightly abuse the terminology and simply refer to 
$\sigma^{\mu\nu}$ as gravito-conductivity). 
The ratio between $\kappa^{\mu\nu}$ and $\sigma^{\mu\nu}$ may be 
defined as the gravitational Seebeck coefficient.

Due to phenomenological considerations, it may be helpful to decompose the 
particle flux \eqref{Transeq} into longitudinal and lateral parts with respect to
the natural orthogonal frame carried by the stationary observer. 
For better understanding, let us rewrite eq.\eqref{Transeq} in the 
orthonormal basis $\{e_{\hat \alpha}{}^\mu\}$, 
$\eta_{\hat\alpha\hat\beta}=g_{\mu\nu}e_{\hat\alpha}{}^\mu e_{\hat\beta}{}^\nu$,
where $\eta_{\hat\alpha\hat\beta}$ represents the Minkowski metric in orthogonal basis and 
$e_{\hat 0}{}^\mu \equiv Z^\mu $. In this basis, the symmetric and the antisymmetric 
parts of $\sigma^{\mu\nu}$ become, respectively,
\begin{align*}
\sigma_{\|}^{\hat a \hat b} = \frac{\mathcal A_{d-1}}{d} 
\zeta \mathcal J_{d+1,-1} \delta^{\hat a \hat b}, \qquad
\sigma_{\bot}^{\hat a \hat b} = 2\tau \frac{\mathcal A_{d-1}}{d} 
\zeta \mathcal J_{d+1,-1} B^{(g)\hat a \hat b},
\end{align*}
where $\hat a,\, \hat b=1,2,\cdots, d$. Naturally, $\sigma_{\|}^{\hat a \hat b}$ 
and $\sigma_{\bot}^{\hat a \hat b}$ obey 
Onsager's reciprocal relations and, respectively, 
characterize the longitudinal ({\it or, diagonal~}) and lateral 
({\it i.e. Hall~}) gravito-conductivities. Let us stress once again that the 
gravito-conductivity presented as above is independent of the 
choice of background geometry and is valid in any stationary spacetimes. 
The only effect of the choice of background geometry is carried by the 
redshifts of the relativistic coldness and the chemical potential 
through Tolman-Ehrenfest and Klein effects.

\section{Example cases and weak field limit}

\subsection{Example cases}

As concrete example cases, let us analyze the gravitational 
conductivity under two nontrivial spacetime backgrounds.
The first example case is the $(d+1)$-dimensional Rindler spacetime with
line element
\begin{align}
\rd  s^2 = g_{\mu\nu}\rd x^\mu \rd x^\nu
= - \kappa^2 x^2 \rd t^2 + \rd x^2
+ \rd \Sigma^2_{d-1},
\label{rindler}
\end{align}
where $\rd \Sigma^2_{d-1}$ denotes the line element of $(d-1)$-dimensional
Euclidean space. The above Rindler spacetime contains an accelerating horizon 
located at $x_h = 0$. 
For a gaseous system of massive particles residing in this accelerating spacetime, 
the coldness will be shifted differently at different spatial places due to 
Tolman-Ehrenfest effect, i.e. $\zeta \sim \kappa x$. In the near horizon limit, we have 
$\zeta \to 0$ and asymptotically $\mathcal J_{d+1,-1} \sim \zeta^{-d}$
(see appendix \ref{C}).
Therefore, we conclude that the longitudinal part of the 
gravito-conductivity tensor $\sigma_{\|}^{\hat a \hat b}$
tends to be divergent in the near horizon limit provided $d>2$, 
and it converges only when the spacetime dimension is $1+1$.

\begin{figure}[h]
	\begin{center}
		\includegraphics[width=.95\textwidth]{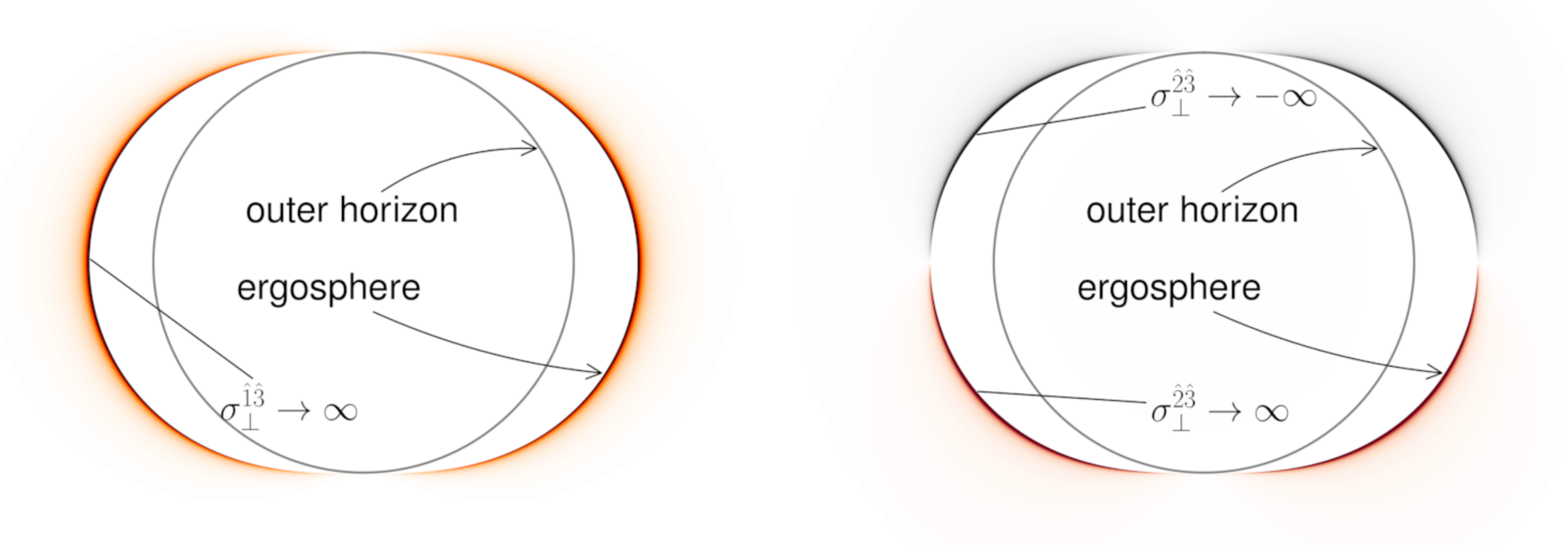}	
		\caption{$\sigma_{\bot}^{\hat 1 \hat 3}$ and $\sigma_{\bot}^{\hat 2 \hat 3}$~:
		darker colors represent bigger amplitudes}
		\label{fig1}
	\end{center}
\end{figure}

In the above example, the GM field defined in eq.\eqref{GEGB} and 
consequently the lateral part of the conductivity tensor is identically 
zero. In order to demonstrate the lateral part, let
us take the Kerr black hole spacetime as the background.
For a gaseous system co-rotating with the spacetime, the coldness 
will be shifted as
\[
\zeta \sim \sqrt{1-\dfrac{r_gr}{r^2+a^2\cos^2\theta}},
\]
where $r_g$ is gravitational radius and $a$ is the Kerr parameter.
In this case $\zeta$ tends to be zero at the
ergosphere, and in this limit we have asymptotically
$\mathcal J_{4,-1}(\alpha,\zeta,\varsigma) \sim \zeta^{-3}$.
As a result, the longitudinal conductivity tensor tends to be divergent at the
ergosphere. For the same reason, the lateral (antisymmetric) part 
of the conductivity tensor also tends to be divergent at the ergosphere. 
The behaviors of $\sigma_{\bot}^{\hat 1 \hat 3}$ and $\sigma_{\bot}^{\hat 2 \hat 3}$
in the near horizon limit are graphically depicted in Fig.\ref{fig1}.

\subsection{Weak field limit}

As an alternative justification of the results presented in the last section, 
let us now consider the weak field limit. In this limit, the stationary observer 
automatically falls back to the static observer in the Minkowski background.

For simplicity, we will consider only the 
cases with $\mathcal D_\nu T = 0$, under which condition the 
covariant transport equation becomes
\begin{align}
j^\mu = - {\tau}{\g}m^{d-1}
\frac{\mathcal A_{d-1}}{d} \zeta \mathcal J_{d+1,-1} \left(\Delta^{\mu\nu} 
+ 2 \tau B^{(g)\mu\nu}\right) \mathcal D_\nu \mu.
\end{align}
Multiplying $\Delta_{\mu\nu} - 2 \tau B^{(g)}_{\mu\nu}$ on both hand sides and 
neglecting terms of order  ${\mathcal O}(\tau^3)$, we get
\begin{align}
j^\mu \left(\Delta_{\mu\nu} - 2 \tau B^{(g)}_{\mu\nu}\right)
=  - {\tau}{\g}m^{d-1}
\frac{\mathcal A_{d-1}}{d} \zeta \mathcal J_{d+1,-1}
\Delta_\nu{}^\alpha \mathcal D_\alpha \mu = - \frac{\tau}{m}
\frac{n^{(0)}}{d} \frac{\zeta \mathcal J_{d+1,-1}}{J_{d-1,1}} 
\Delta_\nu{}^\alpha \mathcal D_\alpha \mu ,
\end{align}
where, in the last step, eq.\eqref{n0} has been used.

Moreover, if $\nabla_\nu \mu = 0$, the 
generalized chemical potential gradient becomes proportional to the 
pure GE field, i.e. 
$\mathcal D_\nu \mu = - \mu E^{(g)}_\nu$. In this case, the transport equation can 
be \blue{rearranged} in the form 
\begin{align}
\mathcal R^{(g)}j_\nu - \mathcal R^{(g)}_H j^\mu B^{(g)}_{\mu\nu} = E^{(g)}_\nu,
\label{nonrellimit}
\end{align}
which looks very similar to eq.\eqref{Hall}, 
wherein gravitational resistance and the gravitational Hall coefficient 
are respectively
\begin{align*}
\mathcal R^{(g)} = \frac{d\,J_{d-1,1}}{\zeta \mathcal J_{d+1,-1}}
\frac{m}{\mu} \frac{1}{n^{(0)}\tau},   \qquad
\mathcal R^{(g)}_H = \frac{d\, J_{d-1,1}}{\zeta \mathcal J_{d+1,-1}}
\frac{m}{\mu} \frac{2}{n^{(0)}}.
\end{align*}
For $d=3$ and taking the non-relativistic limit $\zeta \gg 1$, we have ${3\,J_{2,1}} 
\approx \zeta \mathcal J_{4,-1}$,
$\mu \approx m$. In this case, eq.\eqref{nonrellimit} becomes precisely eq.\eqref{Hall}, 
with
\begin{align*}
\mathcal R^{(g)} \approx \frac{1}{n^{(0)}\tau},   \qquad
\mathcal R^{(g)}_H \approx \frac{2}{n^{(0)}},
\end{align*}
which recovers the results mentioned near the end of Section 2.

\section{Summary}

Transport phenomena under the influence of gravity are among the most 
important macroscopic effects which are relevant to the formation and 
development of astrophysical structures. In the presence of 
gravito-electromagnetism, both longitudinal and lateral transports could arise. 
In the weak field regime, one can adopt the post-Newtonian approximation and 
make use of the the Drude model to describe the gravito-transport 
phenomena, and in such cases the lateral gravito-transport effect could be 
negligibly small. However, in the strong field regime, 
the lateral transport effect may be non-negligible, and even likely to be 
dominated. Therefore, a fully covariant approach for the gravito-transport 
phenomena is needed. 

In this work, we provided a fully covariant approach for the gravito-transport 
equation in the framework of relativistic kinetic theory. The key step is the
introduction of a novel collision model which is different from the 
well-known Anderson-Witting model and  
greatly facilitated the simplification of the relativistic Boltzmann equation. 
In this construction, the proper velocity of the stationary observer played 
an important role. Taking advantage of the novel collision model, 
an approximate near detailed balance solution \eqref{deltaf} for the 
relativistic Boltzmann equation is obtained
for a system of massive neutral particles 
moving in a generic stationary spacetime. 

As an application of \blue{the} solution \eqref{deltaf}, we calculated the particle
flux and derived one of the transport equations.
We found that the gravitational conductivity tensor is temperature dependent,
which is in turn affected by the Tolman-Ehrenfest effect.
Finally, we considered some concrete examples. For neutral fluid in
Rindler spacetime, only longitudinal flux exists and the gravito-conductivity is
divergent at the accelerating horizon. While for neutral fluid co-rotating 
with Kerr spacetime, both the longitudinal and the lateral parts of the 
gravito-conductivity are present and they both tend to be divergent 
at the ergosphere. Whether the divergence of the particle flux at the 
infinite redshift surfaces could be considered as a new
kind of superconductivity deserves further study. Even though, it seems that 
the behaviors of macroscopic systems in the presence of gravity at high temperatures
are quite similar to that of ordinary macroscopic systems in the absence of gravity 
at low temperatures. This similarity has also shown up in the recent study
about the thermodynamic behaviors of AdS black holes in the high temperature limit
\cite{BHhightemp}.

As a final justification of the covariant formalism, we also considered the weak field 
limit and find good agreement with the result that follows from the post 
Newtonian approximation with the use of the Drude model.

\section*{Appendices}
\appendix

\section{Solution to equation \eqref{Cxp} \label{A}}

In this section, we present a recursive solution to \eqref{Cxp}.
In order to find the modification $\delta f$, let us first
decompose the LHS of the RBE into
\begin{align}
\mathscr L_{\mathcal H} f &= p^\sigma\partial_\sigma f
- \frac{1}{2} g^{\mu\sigma} p^\alpha p^\beta \partial_\alpha g_{\beta \sigma}
\frac{\partial f}{\partial p^\mu}
+ m F^{(p)\mu}{}_\nu\, p^\nu \frac{\partial f}{\partial p^\mu}.
\label{LHS}
\end{align}
where $F^{(p)}_{\mu\nu}$ is defined as
\[
F^{(p)}_{\mu\nu} = \frac{1}{m} \nabla_{[\mu}\,p_{\nu]} =
\frac{1}{2m}(\nabla_\mu p_\nu -\nabla_\nu p_\mu).
\]
It is customary to decompose the tensor $F^{(p)}_{\mu\nu}$ into its
``electric'' and ``magnetic'' parts (as observed by the observer $Z^\mu$),
\begin{align}
F^{(p)}_{\mu\nu} = \big(Z_\mu E^{(p)}_\nu-E^{(p)}_\mu Z_\nu\big)
+  B^{(p)}_{\mu\nu},
\end{align}
where
\begin{align}
E^{(p)}_{\mu}= F^{(p)}_{\mu\nu} Z^\nu,
\qquad B^{(p)}_{\mu\nu} = F^{(p)}_{\alpha\beta}
\Delta^\alpha{}_\mu \Delta^\beta{}_\nu.
\end{align}
Let us remind that $B^{(p)}_{\mu\nu}$ should not be confused with 
$B^{(g)}_{\mu\nu}$ defined in eq.\eqref{GEGB}. 
Inserting the collision model \eqref{Cxp} into the RHS of the RBE, we get
\begin{align}
f &= f^{(0)} - \frac{\tau}{\varepsilon}
\mathscr L_{\mathcal H} f.
\label{solu0}
\end{align}
The approximate solution to eq.\eqref{solu0} can be obtained through a recursive 
procedure. At the first order in $\tau$, we can write
\begin{align}
f &\approx f^{(0)} - \frac{\tau}{\varepsilon}
\mathscr L_{\mathcal H} f^{(0)}
= f^{(0)} + \frac{\tau}{\varepsilon} p^\nu \mathscr F^{(0)}_\nu
\frac{\partial f^{(0)}}{\partial \varepsilon},
\label{solu1}
\end{align}
where
\begin{align}
\mathscr F^{(0)}_\nu
%&=\partial_\nu \varepsilon - \partial_\nu \mu
%-\frac{\varepsilon-\mu}{T}\partial_\nu T
%+ \frac{1}{2} Z^\mu p^\sigma \partial_\sigma g_{\mu \nu}- m E^{(g)}_{\nu}  \nonumber \\
%&= \left(\nabla_\nu \mu + \frac{\mu}{c^2}Z^\sigma\nabla_\sigma Z_\nu\right)
%+ \frac{\varepsilon-\mu}{T}\left(\nabla_\nu T+\frac{T}{c^2}Z^\sigma\nabla_\sigma Z_\nu\right)
&= \mathcal D_\nu \mu + \frac{\varepsilon-\mu}{T} \mathcal D_\nu T.
\label{F0}
\end{align}
At this level of the approximation, the magnetic part $B^{(p)}_{\mu\nu}$ of
the tensor $F^{(p)}_{\mu\nu}$ makes no contribution, 
because $Z^\mu\Delta_{\mu\nu}$ is identically zero.
In order to get a refined approximate solution which encodes the effects of the magnetic parts,
we must {\it not} replace all occurrences of $f$ in the 
expression $\mathscr L_\mathcal H f$
by $f^{(0)}$. In order to minimize the required operations, 
we simply keep the distribution function in the last term of eq.\eqref{LHS},
i.e. the magnetic-related terms in its full form. The resulting approximate RBE
becomes
\begin{align}
f = f^{(0)} - \frac{\tau}{\varepsilon} p^\nu \left[
\mathscr F^{(0)}_\nu \frac{\partial f^{(0)}}{\partial \varepsilon}
+ m B^{(p)}{}^\mu{}_\nu
\frac{\partial f}{\partial p^\mu}\right].
\label{solu2}
\end{align}
This is not yet the refined solution because $f$ itself appears in differential form
on the RHS.
To avoid all the complications of solving partial differential equations,
we assume that the final solution still takes a form similar to eq. \eqref{solu1}
but with the effective force $\mathscr F^{(0)}_\mu$ replaced by $\mathscr F_\mu$
which encodes the magnetic contribution:
\begin{align}
f = f^{(0)} + \frac{\tau}{\varepsilon}
p^\nu \mathscr F_\nu \Big(\frac{\partial f^{(0)}}{\partial \varepsilon}\Big).
\label{solu3}
\end{align}
Inserting eq.\eqref{solu3} into the right hand side of eq.\eqref{solu2} 
and comparing the result with eq.\eqref{solu3}, we get the
following algebraic equations for $\mathscr F_\mu$
\begin{align}
\mathscr F_\mu - \frac{\tau m}{\varepsilon}
B^{(p)\nu}_{~\mu} \mathscr F_\nu = \mathscr F^{(0)}_\mu.
\label{solu5}
\end{align}
Following the property of antisymmetric tensor
$ B^{(p)\alpha}_{~\mu} B^{(p)\beta}_{~\alpha} B^{(p)\nu}_{~\beta}
= \dfrac{1}{2}(B^{(p)\beta}_{~\alpha} B^{(p)\alpha}_{~\beta}) B^{(p)\nu}_{~\mu}$,
the solution to eq.\eqref{solu5} can be written as 
$\mathscr F_\mu = \Theta_\mu{}^\nu \mathscr F^{(0)}_\nu$
where the coefficient $\Theta_\mu{}^\nu$ is given by
\begin{align}
\Theta_\mu{}^\nu = \delta_\mu{}^\nu+
\left(1 - \frac{\tau^2 m^2 }{2\varepsilon^2}B^{(p)\beta}_{~\alpha} B^{(p)\alpha}_{~\beta}\right)^{-1}
\left(\frac{\tau m}{\varepsilon} B^{(p)\nu}_{~\mu} + \frac{\tau^2 m^2 }{\varepsilon^2}
B^{(p)\sigma}_{~\mu} B^{(p)\nu}_{~\sigma}\right).
\label{Theta}
\end{align}
Finally the desired solution is found to be $f \approx f^{(0)} + \delta f$, where
the modification $\delta f$ reads
\begin{align}
\delta f
= \frac{\tau}{\varepsilon}\frac{\partial f^{(0)}}{\partial \varepsilon}
p^\mu \, \Theta_\mu{}^\nu \left(\mathcal D_\nu \mu
+ \frac{\varepsilon-\mu}{T} \mathcal D_\nu T\right). \label{solu6}
\end{align}
Truncating the above result at the order $\mathcal O(\tau^2)$, 
eq.\eqref{deltaf} follows.

\section{Derivation of the transport equation \eqref{Transeq} \label{B}}

According to the solution we have just worked out, the desired perturbation
of the particle number flow can be rearranged into the following form
\begin{equation}
\delta N^\nu=\xi^{\mu\nu}\mathcal D_\nu\mu+\chi^{\mu\nu}\mathcal D_\nu T,
\end{equation}
in which the kinetic coefficients are defined as
\begin{equation}
\xi^{\mu\nu}=\la p^\mu p^\sigma\Theta_\sigma{}^\nu\ra,\quad \chi^{\mu\nu}=\frac1T\la(\varepsilon-\mu)p^\mu p^\sigma\Theta_\sigma{}^\nu\ra,
\end{equation}
with the symbol $\la A(p^\mu)\ra$ defined as
\begin{equation*}
\la A(p^\mu)\ra \equiv \int\bm\varpi\frac{\tau}{\varepsilon}\frac{\partial f_0}{\partial \varepsilon}A(p^\mu).
\end{equation*}
Please note that $\Theta_\mu{}^\nu$ can be expanded in powers of $\tau$,
\begin{equation*}
\Theta_\mu{}^\nu\approx \delta_\mu{}^\nu+\frac{\tau m}{\varepsilon}B^{(p)}_\mu{}^\nu+\frac{\tau^2 m^2}{\varepsilon^2}B^{(p)}_\mu{}^\sigma B^{(p)}_\sigma{}^\nu
+\cdots.
\end{equation*}
Therefore, taking advantage of the orthonormal basis $\{e_{\hat\alpha}{}^\mu\}$, 
the momentum can be expanded as $p^\mu=p^{\hat\alpha}(e_{\hat\alpha})^\mu$, 
and it follows that
\begin{align}
\xi^{\mu\nu}&=[\xi^{(1)}]^{\mu\nu}+[\xi^{(2)}]^{\mu\nu},
\end{align}
where
\begin{align}
\left[\xi^{(1)}\right]^{\mu\nu}
&=-\g \zeta \tau\mathcal A_{d-1}{m}^{d-1}
\left[\mathcal J_{d-1,1} \, {Z^\mu Z^\nu}
+\frac{1}{d} \mathcal J_{d+1,-1}
\, \Delta^{\mu\nu} \right],\\
[\xi^{(2)}]^{\mu\nu}&=m\left\la\frac{\tau}{\varepsilon}p^\mu p^\sigma B^{(p)}_\sigma{}^\nu\right\ra
=m\int\bm\varpi\frac{\tau^2}{\varepsilon^2}\frac{\partial f_0}{\partial \varepsilon}p^\mu p^\sigma B^{(p)}_\sigma{}^\nu\notag\\
&=-\frac{1}{2}\g\zeta\tau^2\mathcal{A}_{d-1}{m}^{d-1}\mathcal{J}_{d-1,1}e_{\hat0}{}^\mu e_{\hat0}{}^\sigma e_{\hat0}{}^\rho (\partial_\alpha g_{\beta\rho}-\partial_\beta g_{\alpha\rho})\Delta^\alpha{}_\sigma\Delta^{\beta\nu}\notag\\
&\quad-\frac{\g\zeta\tau^2\mathcal{A}_{d-1}m^{d-1}}{2d}\mathcal{J}_{d+1,-1}(\delta^{\hat a\hat b}e_{\hat a}{}^\mu e_{\hat b}{}^\sigma e_{\hat0}{}^\rho+\delta^{\hat a\hat c}e_{\hat a}{}^\mu e_{\hat 0}{}^\sigma e_{\hat c}{}^\rho+\delta^{\hat b\hat c}e_{\hat 0}{}^\mu e_{\hat b}{}^\sigma e_{\hat c}{}^\rho)\notag\\
&\quad\times(\partial_\alpha g_{\beta\rho}-\partial_\beta g_{\alpha\rho})\Delta^\alpha{}_\sigma\Delta^{\beta\nu}.
\end{align}
By identifying the proper velocity $Z^\mu$ of the stationary observer with 
$(e_{\hat0}{}^\mu)$ and recalling the identity $Z^\mu \Delta_{\mu\nu}=0$, 
the last equation can be further simplified, yielding
\begin{equation*}
[\xi^{(2)}]^{\mu\nu}=-\frac{\g\zeta\tau^2\mathcal{A}_{d-1}m^{d-1}}{2d}
\mathcal{J}_{d+1,-1}(Z^\rho\Delta^{\mu\alpha}+Z^\mu\Delta^{\rho\alpha})(\partial_\alpha g_{\beta\rho}-\partial_\beta g_{\alpha\rho})\Delta^{\beta\nu}.
\end{equation*}
Moreover, since the GM field $B^{(g)}_{\mu\nu}$ obeys the relation
\begin{equation*}
B^{(g)}_{\mu\nu}=\frac{1}{2}\Delta_{[\mu}{}^\alpha\Delta_{\nu]}{}^\beta\nabla_\alpha Z_\beta,
\end{equation*}
$[\xi^{(2)}]^{\mu\nu}$ can be further rewritten as
\begin{align}
[\xi^{(2)}]^{\mu\nu}=&-\frac{\g\zeta\tau^2\mathcal{A}_{d-1}m^{d-1}}{d}
\mathcal{J}_{d+1,-1}(2B^{(g)\mu\nu}+Z^\mu\partial_{[\alpha}g_{\beta]\rho}\Delta^{\rho\alpha}\Delta^{\beta\nu}).
\end{align}
Similarly, we have
\begin{align}
\chi^{\mu\nu}&=\frac{1}{T}\la(\varepsilon-\mu)p^\mu p^\sigma\Theta_\sigma{}^\nu\ra\notag\\
&=\frac{1}{T}\la (\varepsilon-\mu)p^\mu p^\nu\ra+\frac1T\la(\varepsilon-\mu)
\frac{\tau m}{\varepsilon}p^\mu p^\sigma B^{(p)}_\sigma{}^\nu\ra\notag\\
    &=[\chi^{(1)}]^{\mu\nu}+[\chi^{(2)}]^{\mu\nu},
\end{align}
where at the leading order,
\begin{align}
[\chi^{(1)}]^{\mu\nu}
   =\frac{\g\zeta\tau\mathcal{A}_{d-1}m^{d-1}}{T}
   \bigg[(\mu\mathcal{J}_{d-1,1} &-m \mathcal{J}_{d-1,2}) {Z^\mu Z^\nu} \notag\\
   &+\frac1d(\mu\mathcal{J}_{d+1,-1}-m \mathcal{J}_{d+1,0})\Delta^{\mu\nu}\bigg],
\end{align}
and at the next to leading order,
\begin{equation}
[\chi^{(2)}]^{\mu\nu}=\frac{\g\zeta\tau^2\mathcal{A}_{d-1}m^{d-1}}{dT}(2B^{(g)\mu\nu}
+Z^\mu\partial_{[\alpha}g_{\beta]\rho}\Delta^{\alpha\rho}\Delta^{\beta\nu})(\mu\mathcal{J}_{d+1,-1}-m \mathcal{J}_{d+1,0}).
\end{equation}
Combining together, we have
\begin{align}
\delta N^\mu&=\xi^{\mu\nu}\mathcal{D}_\nu\mu+\chi^{\mu\nu}\mathcal{D}_\nu T\notag\\
%&=([\sigma^{(1)}]^{\mu\nu}+[\sigma^{(2)}]^{\mu\nu})\mathcal{D}_\nu\mu+([\beta^{(1)}]^{\mu\nu}+[\beta^{(2)}]^{\mu\nu})\mathcal{D}_\nu T\notag\\
&=\left\{-\g\zeta\tau\mathcal A_{d-1}m^{d-1}
\left[\mathcal J_{d-1,1} \, {Z^\mu Z^\nu}
+\frac{1}{d} \mathcal J_{d+1,-1}
\, \Delta^{\mu\nu} \right]\right.\notag\\
&\left.-\frac{\g\zeta\tau^2\mathcal A_{d-1}m^{d-1}}{d}
\mathcal J_{d+1,-1}(2B^{(g)\mu\nu}+Z^\mu\partial_{[\alpha}g_{\beta]\rho}
\Delta^{\rho\alpha}\Delta^{\beta\nu})\right\}\mathcal{D}_\nu\mu\notag\\
&+\left\{\frac{\g\zeta\tau\mathcal A_{d-1}m^{d-1}}{T}\left[(\mu\mathcal J_{d-1,1}-m\mathcal J_{d-1,2}) {Z^\mu Z^\nu}
+\frac1d(\mu\mathcal J_{d+1,-1}-m\mathcal J_{d+1,0})\Delta^{\mu\nu}\right]\right.\notag\\
&\left.+\frac{\g\zeta\tau^2\mathcal A_{d-1}m^{d-1}}{dT}(2B^{(g)\mu\nu}+Z^\mu\partial_{[\alpha}g_{\beta]\rho}
\Delta^{\alpha\rho}\Delta^{\beta\nu})(\mu\mathcal J_{d+1,-1}-m\mathcal J_{d+1,0})\right\}\mathcal{D}_\nu T.
\label{dNmu}
\end{align}
Finally, taking the orthogonal projection $j^\mu=\Delta^\mu{}_\nu\delta N^\nu$, 
all terms with the factor $Z^\mu$ in eq.\eqref{dNmu} drops off and we are led to
the result
\begin{align*}
j^\mu%=\Delta^\mu{}_\nu\delta N^\nu
=-\frac{\g\zeta\tau\mathcal A_{d-1}m^{d-1}}{d}(2\tau B^{(g)\mu\nu}+\Delta^{\mu\nu})
&\bigg[\mathcal J_{d+1,-1}\mathcal D_\nu\mu %\\
%&
-\frac1T(\mu\mathcal J_{d+1,-1}-m\mathcal J_{d+1,0})\mathcal D_\nu T\bigg].
\end{align*}
With a little more simplification, this result can be rearranged in the 
form \eqref{Transeq}.

\section{About the integral $\mathcal J_{d+1,-1}$ \label{C}}

To illustrate the asymptotic behavior of the integral $\mathcal J_{d+1,-1}$,
we consider the special case $d=3$ 
and take the high temperature limit. For the sake of simplicity we also 
set $\varsigma = 0$. Then the integral $\mathcal J_{4,-1}$ reduces to
\begin{equation}\label{special}
\mathcal J_{4,-1} \sim I=\int_0^\infty\frac{\sinh^4\vartheta}{\cosh\vartheta}
\re^{-\zeta\cosh\vartheta}\rd\vartheta,
\end{equation}
which can be rewritten as
\begin{align}\label{reJ4-1}
I&=\int_0^\infty\left(\cosh^3\vartheta-2\cosh\vartheta+\frac{1}{\cosh\vartheta}\right)\re^{-\zeta\cosh\vartheta}\rd\vartheta\notag\\
%&=\int_0^\infty\left(\frac14\cosh3\vartheta-\frac54\cosh\vartheta+\frac{1}{\cosh\vartheta}\right)\re^{-\zeta\cosh\vartheta}\rd\vartheta\notag\\
&=\frac14K_3(\zeta)-\frac54K_1(\zeta)+\int_0^\infty\frac{\re^{-\zeta\cosh\vartheta}}{\cosh\vartheta}\rd\vartheta,
\end{align}
where $K_n(\zeta)$ is the modified Bessel function of the second kind. 
At the high temperature limit,
$\zeta \ll 1$, we have
\begin{equation}
K_\nu(\zeta)\sim\frac{\Gamma(\nu)}{2\left(\frac12\zeta\right)^\nu},\quad(\nu>0),
\end{equation}
and according to the identity
\begin{equation}
\int_0^\infty\frac{\cos(ax)}{\cosh(\beta x)}\rd x=\frac{\pi}{2\beta}\sech\left(\frac{\pi a}{2\beta}\right),\quad \rm{Re}\beta>0,\quad a\in\mathbb{R},
\end{equation}
the last term of \eqref{reJ4-1} is convergent. Therefore, at the leading order, we have
\begin{equation}
\mathcal J_{4,-1} \sim\frac2{\zeta^3}, \qquad ({\rm as} \,\,\,\zeta \to 0).
\end{equation}
For arbitrary choices of the spacetime dimensions and degeneracy 
parameter $\varsigma$, the asymptotic behavior of the integral 
$\mathcal J_{d+1,-1}$ is always $\mathcal J_{d+1,-1} \propto \zeta^{-d}$
as $\zeta\to 0$, which can be verified numerically.

\section*{Acknowledgement}

X. H.  is supported by the Hebei NSF under grant No. A2021205037. 
L.Z. is supported by the National Natural Science Foundation of China 
under the grant No. 12275138.

\section*{Data Availability Statement} 

This manuscript has no associated data. 

\section*{Declaration of competing interest}

The authors declare no competing interest.

\providecommand{\href}[2]{#2}\begingroup%\raggedright
\footnotesize\itemsep=0pt
\providecommand{\eprint}[2][]{\href{http://arxiv.org/abs/#2}{arXiv:#2}}

\endgroup


\begin{thebibliography}{10}

\bibitem{EventHorizonTelescope:2019dse}
K.~Akiyama \textit{et al.} [Event Horizon Telescope],
``First M87 event horizon telescope results. I. The shadow of the supermassive black hole,''
\href{https://iopscience.iop.org/article/10.3847/2041-8213/ab0ec7}{Astrophys. J. Lett. \textbf{875}, L1 (2019)}.

\bibitem{doi:10.1002/andp.19113390503}
F.~J\"uttner, ``Das maxwellsche gesetz der geschwindigkeitsverteilung in
  der relativtheorie,''
  \href{https://doi.org/10.1002/andp.19113390503}{Annalen der Physik
  {\bfseries 339} 856--882}.

\bibitem{Synge:1934zzb}
J.~L. Synge, ``The energy tensor of a continuous medium,'' {Trans.
  Roy. Soc. Canada {\bfseries III 28} (1934) 127}.

\bibitem{PhysRev.122.1342}
G.~E. Tauber and J.~W. Weinberg, ``Internal state of a gravitating gas,''
  \href{https://doi.org/10.1103/PhysRev.122.1342}{Phys. Rev. {\bfseries122} (May, 1961) 1342--1365}.

\bibitem{Chernikov1963}
N.~A. Chernikov, ``The relativistic gas in the gravitational field,''
  {Acta Phys.Polon. {\bfseries 23 (1963)} 629}.

\bibitem{doi:10.1063/1.1704047}
W.~Israel, ``Relativistic kinetic theory of a simple gas,''
  \href{https://doi.org/10.1063/1.1704047}{J. Math. Phys. {\bfseries 4} (1963) 1163--1181}.

\bibitem{Heinz:1983nx}
U.~W.~Heinz,
``Kinetic Theory for Nonabelian Plasmas,''
\href{https://doi.org/10.1103/PhysRevLett.51.351}{Phys. Rev. Lett. \textbf{51}, 351 (1983)}.

\bibitem{Bazow:2015dha}
D.~Bazow, G.~S.~Denicol, U.~Heinz, M.~Martinez and J.~Noronha,
``Analytic solution of the Boltzmann equation in an expanding system,''
\href{https://doi.org/10.1103/PhysRevLett.116.022301}{Phys. Rev. Lett. \textbf{116}, no.2, 022301 (2016)}.

\bibitem{Rioseco:2016jwc}
P.~Rioseco and O.~Sarbach,
``Accretion of a relativistic, collisionless kinetic gas into a Schwarzschild black hole,''
\href{https://doi.org/10.1088/1361-6382/aa65fa}{Class. Quant. Grav. \textbf{34}, no.9, 095007 (2017)}.

\bibitem{Guo-Qiang:1996}
C.~M.~Ko and G.~Q.~Li,
``Medium effects in high-energy heavy ion collisions,''
\href{https://doi.org/10.1088/0954-3899/22/12/002}{J.Phys.G 22 (1996) 1673-1726}.

\bibitem{Miake:2005}
K.~Yagi, T.~Hatsuda and Y.~Miake,
``Quark-gluon Plasma,''
\href{https://doi.org/10.1080/00107510902978246}{Cambridge University Press, 2005}.

\bibitem{Mendoza:2014}
S.~Sauro et al,
``Relativistic lattice kinetic theory: Recent developments and future prospects.''
\href{https://doi.org/10.1140/epjst/e2014-02257-0}{Eur. Phys. J. Spec. Top. 223, 2177¨C2188 (2014)}.

\bibitem{Greif:2014oia}
M.~Greif, I.~Bouras, C.~Greiner and Z.~Xu,
``Electric conductivity of the quark-gluon plasma investigated using a perturbative QCD based parton cascade,''
\href{https://doi.org/10.1103/PhysRevD.90.094014}{Phys. Rev. D 90, 094014 (2014)}
[\eprint{1408.7049}].

\bibitem{Simeoni:2019}
A.~Gabbana, D.~Simeoni, S.~Succi and R.~Tripiccione,
``Relativistic lattice Boltzmann methods: Theory and applications,''
\href{https://doi.org/10.1016/j.physrep.2020.03.004}{Phys. Rept. \textbf{863} (2020), 1-63}
[\eprint{1909.04502}].

\bibitem{ChunShen:2022}
K.~Sun, R.~Wang, et al,
``Unveiling the dynamics of nucleosynthesis in relativistic heavy-ion collisions,''
[\eprint{2207.12532}].

\bibitem{Elze:1987ii}
H.~T.~Elze, M.~Gyulassy, D.~Vasak, H.~Heinz, H.~Stoecker and W.~Greiner,
``Towards a relativistic self consistent quantum transport theory of hadronic matter,''
\href{https://doi.org/10.1142/S0217732387000562}{Mod. Phys. Lett. A \textbf{2} (1987), 451-460}.

\bibitem{Danielewicz:1991dh}
P.~Danielewicz and G.~F.~Bertsch,
``Production of deuterons and pions in a transport model of energetic heavy ion reactions,''
\href{https://doi.org/10.1016/0375-9474(91)90541-D}{Nuclear Physics A, 1991, 533(4):712-748}.

\bibitem{Bertsch:1988}
G.~F.~Bertsch and S.~Das~Gupta,
``A guide to microscopic models for intermediate energy heavy ion collisions,''
\href{https://doi.org/10.1016/0370-1573(88)90170-6}{Physics Reports, 1997, 160(4):189-233}.

\bibitem{TMEP:2022xjg}
H.~Wolter \textit{et al.},
``Transport model comparison studies of intermediate-energy heavy-ion collisions,''
\href{https://doi.org/10.1016/j.ppnp.2022.103962}{Prog. Part. Nucl. Phys. 125 (2022) 103962}
[\eprint{2202.06672}].



\bibitem{Hannestad:1995rs}
S.~Hannestad and J.~Madsen,
``Neutrino decoupling in the early universe,''
\href{https://doi.org/10.1103/PhysRevD.52.1764}
{Phys. Rev. D \textbf{52} (1995), 1764-1769}
[\eprint{astro-ph/9506015}].

\bibitem{Hu:1995em}
W.~T.~Hu,
``Wandering in the background: A CMB explorer,''
[\eprint{astro-ph/9508126}].

\bibitem{Uzan:1998mc}
J.~P.~Uzan,
``Dynamics of relativistic interacting gases: From a kinetic to a fluid description,''
\href{https://doi.org/10.1088/0264-9381/15/4/025}
{Class. Quant. Grav. \textbf{15} (1998), 1063-1088}
[\eprint{gr-qc/9801108}].

\bibitem{Birrell:2014gea}
J.~Birrell, J.~Wilkening and J.~Rafelski,
``Boltzmann equation solver adapted to emergent chemical non-equilibrium,''
\href{https://doi.org/10.1016/j.jcp.2014.10.056}{J. Comput. Phys. \textbf{281} (2015), 896-916}
[\eprint{1403.2019}].

\bibitem{Husdal:2016pfd}
L.~Husdal and I.~Brevik,
``Entropy production in a lepton-photon universe,''
\href{https://doi.org/10.1007/s10509-017-3023-1}{Astrophys. Space Sci. \textbf{362} (2017) no.2, 39}
[\eprint{1610.04451}].

\bibitem{Adshead:2016xxj}
P.~Adshead, Y.~Cui and J.~Shelton,
``Chilly dark sectors and asymmetric reheating,''
\href{https://doi.org/abs/10.1007/JHEP06(2016)016}{JHEP \textbf{06} (2016), 016}
[\eprint{1604.02458}].

\bibitem{Sasankan:2019oee}
N.~Sasankan, A.~Kedia, M.~Kusakabe and G.~J.~Mathews,
``Analysis of the multi-component relativistic Boltzmann equation for electron scattering in big bang nucleosynthesis,''
\href{doi:10.1103/PhysRevD.101.123532}
{Phys. Rev. D 101, 123532 (2020)}
[\eprint{1911.07334}].

\bibitem{Pordeus-da-Silva:2019bak}
G.~Pordeus-da-Silva, R.~C.~Batista and L.~G.~Medeiros,
``Theoretical foundations of the reduced relativistic gas in the cosmological perturbed context,''
\href{doi:10.1088/1475-7516/2019/06/043}{JCAP \textbf{06} (2019), 043}
[\eprint{1904.09904}].

\bibitem{Pitrou:2019hqg}
C.~Pitrou,
``Radiative transport of relativistic species in cosmology,''
\href{https://doi.org/abs/10.1016/j.astropartphys.2020.102494}{Astropart. Phys. \textbf{125} (2021), 102494}
[\eprint{1902.09456}].

\bibitem{Sarbach2022}
R.~O.~Acu\~na-C\'ardenas, C.~Gabarrete and O.~Sarbach,
``An introduction to the relativistic kinetic theory on curved spacetimes,''
\href{https://doi.org/10.1007/s10714-022-02908-5}{Gen. Rel. Grav. \textbf{54}, no.3, 23 (2022)}
[\eprint{2106.09235}].



\bibitem{Molnar:2020nfd}
S.~M.~Molnar and J.~Godfrey,
``Empirical test for relativistic kinetic theories based on the Sunyaev\textendash{}Zel\textquoteright{}dovich effect,''
\href{https://doi.org/10.3847/1538-4357/abb6f6}{Astrophys. J. \textbf{902}, no.2, 143 (2020)}

\bibitem{Deng:2021twz}
X.~C.~Deng, W.~Hu, F.~W.~Lu and B.~Z.~Dai,
``Kinetic powers of the relativistic jets in Mrk 421 and Mrk 501,''
\href{https://doi.org/10.1093/mnras/stab919}{Mon. Not. Roy. Astron. Soc. \textbf{504}, no.1, 878-887 (2021)}

\bibitem{Nishikawa:2020rwe}
K.~Nishikawa, I.~Dutan, C.~Koehn and Y.~Mizuno,
``PIC methods in astrophysics: Simulations of relativistic jets and kinetic physics in astrophysical systems,''
\href{https://doi.org/10.1007/s41115-021-00012-0}{Liv. Rev. Comput. Astrophys. \textbf{7}, 1 (2021)}

\bibitem{Mach:2022mrn}
P.~Mach and A.~Odrzywolek,
``Accretion of the relativistic Vlasov gas onto a moving Schwarzschild black hole: Low-temperature limit and numerical aspects,''
\href{https://doi.org/10.5506/APhysPolBSupp.15.1-A7}{Acta Phys. Polon. Supp. \textbf{15}, no.1, 1 (2022)}


\bibitem{Rioseco:2017b}
P.~Rioseco and O.~Sarbach,
``Spherical steady-state accretion of a relativistic
collisionless gas into a Schwarzschild black hole,''
[\eprint{1701.07104}].

\bibitem{Gabarrete:2021}
A.~Gamboa, C.~Gabarrete, P.~Domi­nguez-Fernandez, D.~Nunez
and O.~ Sarbach,
``Accretion of a Vlasov gas on to a black hole from a sphere of finite radius
and the role of angular momentum,''
[\eprint{2107.04830}].

\bibitem{Israel:1976tn}
W.~Israel,
``Nonstationary irreversible thermodynamics: A Causal relativistic theory,''
\href{https://doi.org/10.1016/0003-4916(76)90064-6}{Annals Phys. \textbf{100}, 310-331 (1976)}.

\bibitem{Israel:1979wp}
W.~Israel and J.~M.~Stewart,
``Transient relativistic thermodynamics and kinetic theory,''
\href{https://doi.org/10.1016/0003-4916(79)90130-1}{Annals Phys. \textbf{118}, 341-372 (1979)}.

\bibitem{Anderdon:1973}
J.~L.~Anderdon and H.~R.~Witting
``A realtivistic relaxtion-time model for the Boltzmann equation,''
\href{https://doi.org/10.1016/0031-8914(74)90355-3}{Physica 74 (1974) 466-488}

\bibitem{deGroot}
S.~R.~De Groot, W.~A.~Van Leeuwen and C.~G.~Van Weert,
``Relativistic kinetic theory: Principles and applications,'' (1980)

\bibitem{Cercignani:2002}
C.~Cercignani and G.~Kremer,
  ``{The relativistic Boltzmann equation: theory and applications},''
  \href{https://doi.org/10.1007/978-3-0348-8165-4}{{Birkh{\"a}user Basel}
  (2002)}.

\bibitem{Hao:2021ifw}
X.~Hao, S.~Liu and L.~Zhao,
``Relativistic transformation of thermodynamic parameters and refined Saha equation,''
[\eprint{2105.07313}].

\bibitem{Thorne}
K.~S. Thorne, ``Gravitomagnetism, jets in quasars, and the stanford
  gyroscope experiment, from the book: New frontiers of physics,'' {W. H.
  Freeman and Company, New York, 1988}.

\bibitem{Mashhoon:2000he}
B.~Mashhoon, ``Gravitoelectromagnetis,'' in {Spanish Relativity
  Meeting on Reference Frames and Gravitomagnetism (EREs2000) Valladolid,
  Spain, September 6-9}, 2000,
  [\eprint{gr-qc/0011014}].

\bibitem{DeWitt:1966yi}
B.~S. DeWitt, ``{Superconductors and gravitational drag},''
  \href{https://doi.org/10.1103/PhysRevLett.16.1092}{Phys. Rev. Lett.}
  {\bfseries 16} (1966) 1092--1093.

\bibitem{Dessler:1968zz}
A.~J. Dessler, F.~C. Michel, H.~E. Rorschach and G.~T. Trammell,
  ``{Gravitationally induced electric fields in conductors},''
  \href{https://doi.org/10.1103/PhysRev.168.737}{Phys. Rev. {\bfseries
  168} (1968) 737--743}.
  
\bibitem{Israel}
W.~Israel and J.~M.~Stewart
``Transient relativistic thermodynamics and kinetic theory,''
\href{https://doi.org/10.1016/0003-4916(79)90130-1}{Annals of Physics \textbf{118}, 2, 341-372 (1979)}
  
\bibitem{BGK1954}
P.~L.~Bhatnagar, E.~P.~Gross, and M.~Krook,
``A model for collision processes in gases. I. Small amplitude processes 
in charged and neutral one-component systems,'' 
\href{https://doi.org/10.1103/PhysRev.94.511}{Phys. Rev. {\bfseries 94} (1954) 511--525}.

\bibitem{Marle1965}
C.~Marle,
``Mod{\`e}le cinetique pour l'etablissement des lois de la conduction de la chaleur et de la
viscosite en th{\'e}orie de la relativite,'' 
%\href{https://zbmath.org/?q=an:0127.17408}
{C. R. Acad. Sci., Paris {\bfseries 260} (1965) 6539-6541}.


\bibitem{BHhightemp} X. Kong, T. Wang, L. Zhao, ``High temperature AdS black 
holes are low temperature quantum phonon gases,'' 
 [\eprint{2209.12230}].


\end{thebibliography}
\end{document}